\documentclass[letterpaper]{jpconf}
\usepackage{graphicx}
\usepackage{dcolumn}
\usepackage{bm}
\usepackage{epsfig}
\usepackage{cite}
\usepackage{subfigure}

\begin{document}

\title{Recent Results on Strangeness Production at RHIC}

\author{M A C Lamont\dag for the STAR Collaboration}

\address{\dag\ Physics Department, Yale University, New Haven, CT 06520, USA}
\ead{Matthew.Lamont@yale.edu}

\begin{abstract}

Due to its large acceptance, the STAR experiment has acquired a wealth of data on strangeness production for a variety of collisions systems and energies, from p+p to Au+Au.  By using the yields and spectra, we address the evolution of the bulk system, including strangeness enhancement and the flavour dependence of radial and elliptic flow.  Utilising the fact that we can identify strange baryons and mesons, we investigate different hadronization mechanisms in the intermediate and high p$_{T}$ regions.  The ratios of the particle yields, measured to high p$_{T}$, are used to further investigate the range and applicability of the previously reported anomalous baryon production.  We also study two-particle azimuthal correlations of identified particles in order to investigate any flavour dependence of jet fragmentation in the available p$_{T}$ range.  Data was presented for a number of different collision systems and energies.

\end{abstract}




\section{Introduction}

In this paper, I will present an overview of recent results on strangeness production from the STAR collaboration at RHIC.  The data will be split into two main sections, the first part details data at low p$_{T}$, which drives the bulk observables (such as integrated yields and elliptic flow).  The second part of this paper discusses strangeness production mechanisms at intermediate p$_{T}$ in the form of yields, ratios and two-particle azimuthal correlations.

\section{Strangeness in the Bulk}

\begin{figure}[h!]
\begin{center}
\includegraphics[width=0.4\textwidth, height=0.35\textwidth]{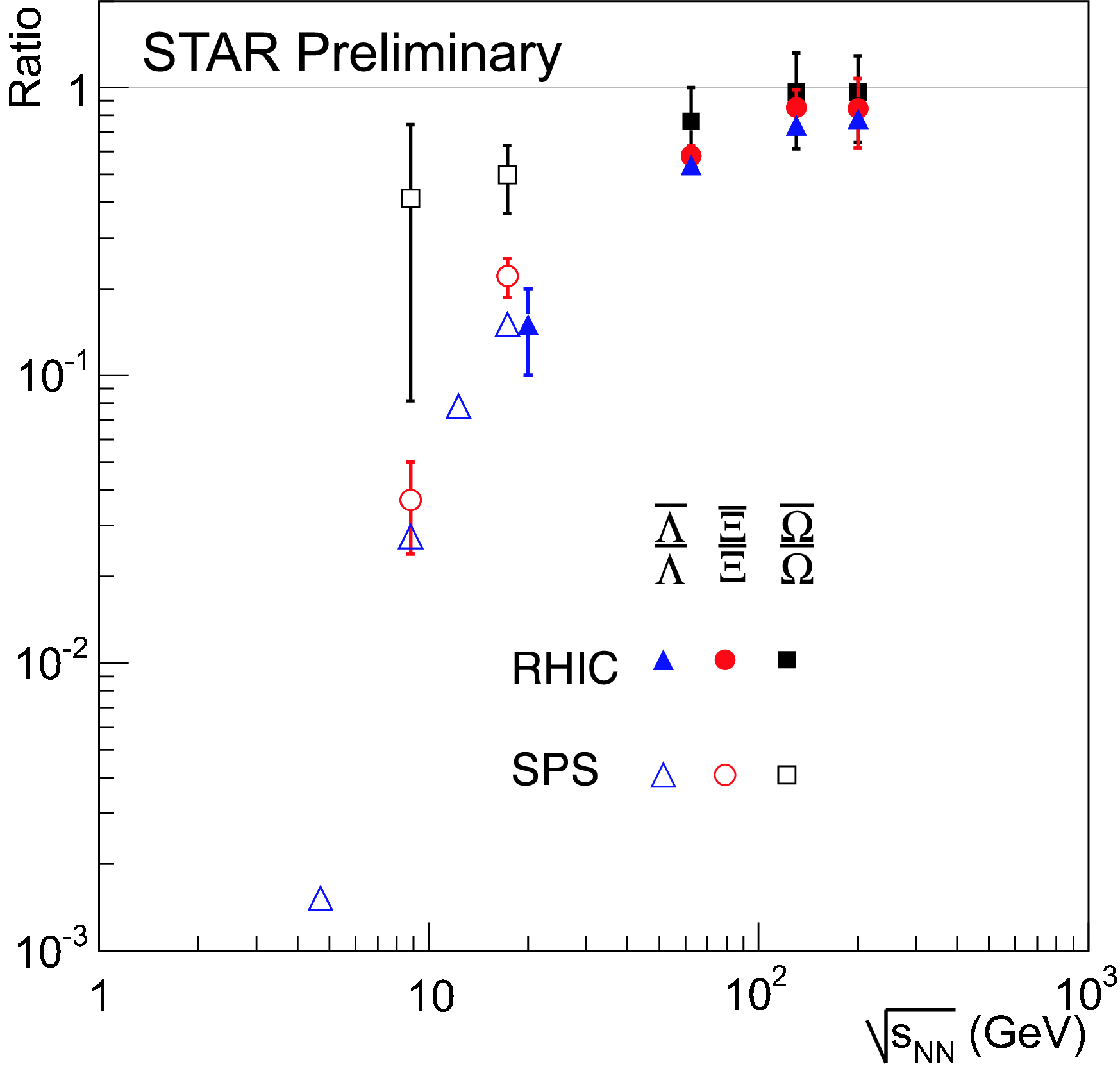}
\end{center}
\caption{Colour online: Strange anti-baryon to baryon ratios in heavy-ion collisions as a function of centre-of-mass energy.}
\label{Fig:Ratios}
\end{figure}

Due to the high statistics data-sets available at RHIC, STAR has collected a large body of data on strange particles.  Using this data, we can compare many observables as a function of system size and energy.  Figure \ref{Fig:Ratios} shows one such observable, the evolution of the strange anti-baryon/baryon ratios as a function of energy in heavy-ion collisions.  It is evident that the evolution of the ratios is smooth with energy, over almost two orders of magnitude.  This in turn is a reflection of the fact that there is a smooth change in the net-baryon density at mid-rapidity in the collision, going from near-zero at the top RHIC energies to large net-baryon densities at SPS energies.  The ratios also exhibit an ordering with strangeness content (where $\overline{\Omega}$/$\Omega > \overline{\Xi}/\Xi > \overline{\Lambda}/\Lambda$). This is as expected as neither the $\Omega$ nor the  $\overline{\Omega}$ share a valence quark with the incoming nucleons and so are not as affected by the net-baryon density as is the $\Lambda$.

\subsection{Strangeness Enhancement at RHIC}

One of the original proposed signatures of de-confinement was that of an enhancement of the production of strange particles in heavy-ion collisions relative to more elementary systems ~\cite{Rafelski}.  This enhancement is due to a lower threshold energy for producing strangeness in a de-confined medium (twice the strange quark mass), together with the extra gluonic channels available for strange quark production.  The enhancement is expected to be smaller for collisions at larger $\sqrt{s}$ and should be greater for particles containing more (anti)strange quarks. 

\begin{figure}[htb]
\centering
\mbox{\subfigure[Colour online: Enhancement factors as a function of collision centrality (determined by $N_{part}$) for AA collisions relative to p+p (STAR) and p+Be (NA57) collisions respectively.]{\includegraphics[width=0.49\textwidth,height=0.4\textwidth]{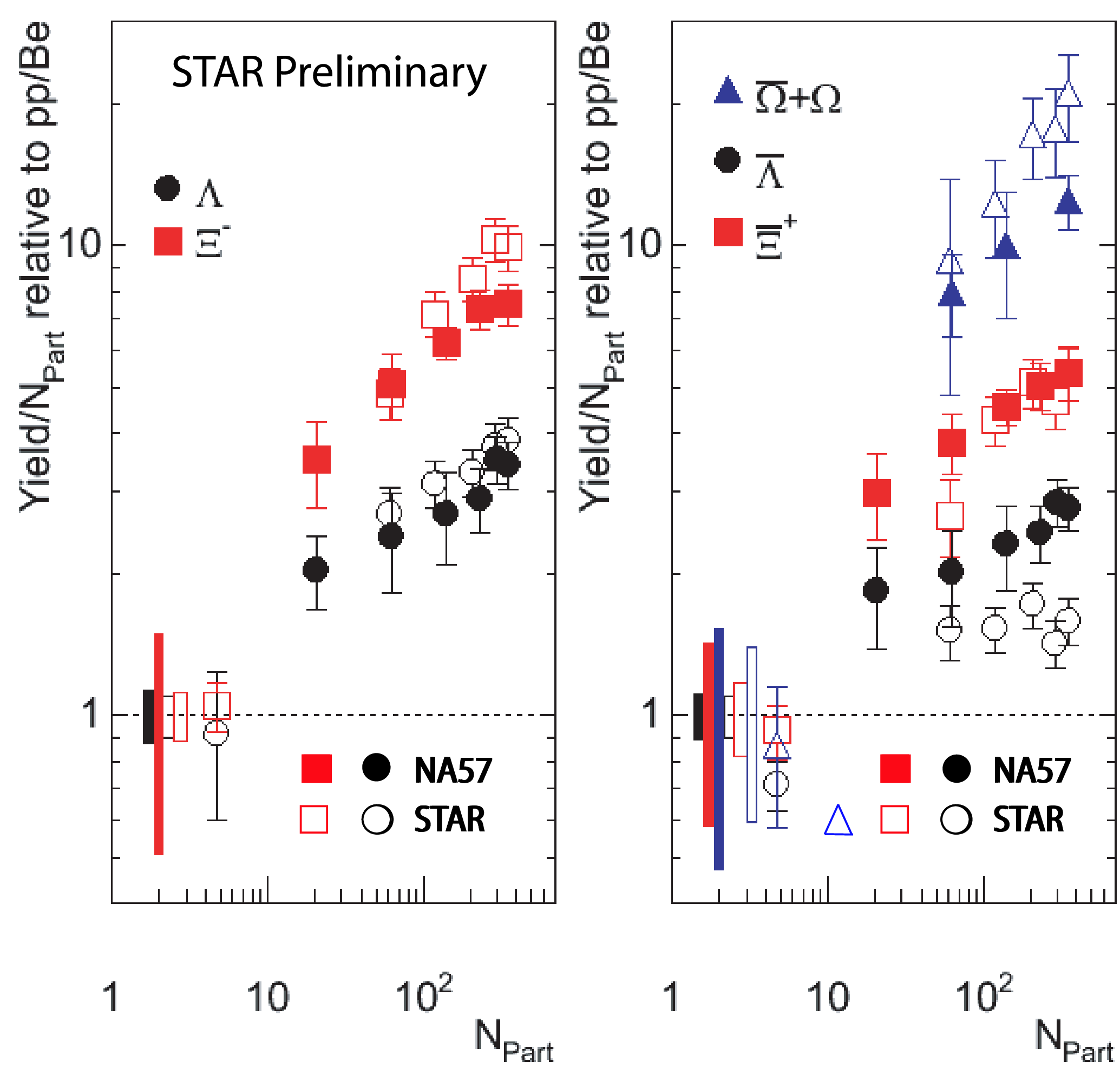}}\quad
\subfigure[Colour online: Theory curves describing evolution of freeze-out parameters for different geometry assumptions (see text for details).]{\includegraphics[width=0.49\textwidth,height=0.4\textwidth]{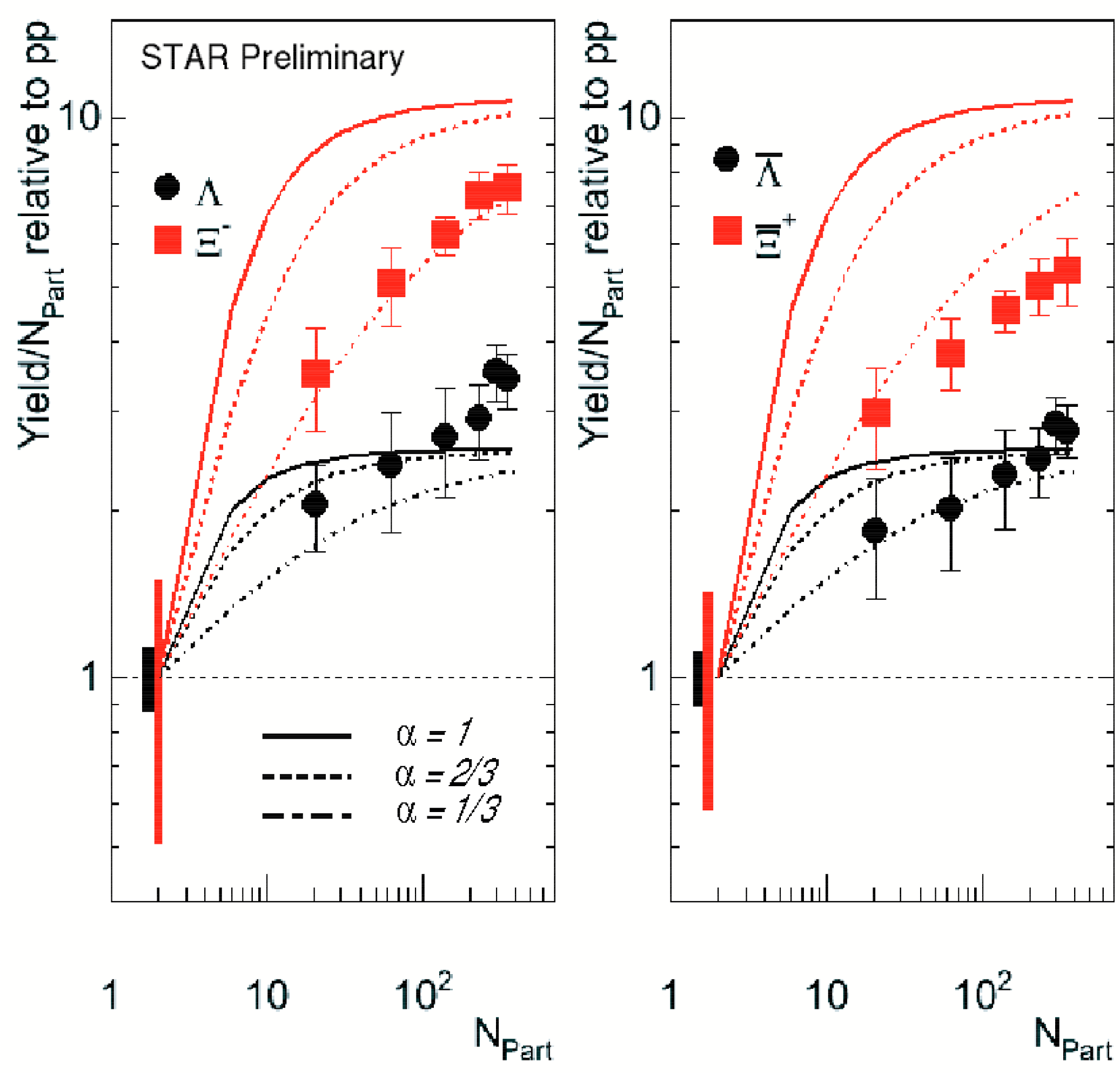}}}
\vspace{-0.5cm}
\caption{}
\label{Fig:EnhanceComp}
\end{figure}

Figure \ref{Fig:EnhanceComp}a shows the enhancement of strange baryons and anti-baryons as a function of centrality as measured by STAR for Au+Au collisions at $\sqrt{s_{NN}}$ = 200 GeV and by NA57 for Pb+Pb collisions at $\sqrt{s_{NN}}$ = 17.3 GeV~\cite{NA57}.  The enhancement of the STAR data is relative to p+p collisions, whilst the NA57 data is relative to p+Be collisions.  The enhancement of the $\overline{\Omega}+\Omega$ from STAR is shown here for the first time.  It is immediately obvious that there is very little difference between the two systems (apart from the case of the $\Lambda$).  This is also the same for lower energy SPS data which is in contradiction to the energy dependence which was predicted~\cite{RedlichEnhance}.  Any difference arising from comparison to p+Be collisions instead of p+p collisions has been shown to be negligible~\cite{NA49}.

%

Figure \ref{Fig:EnhanceComp}b shows the STAR enhancement data only, overlaid with three theory curves~\cite{RedlichCurves}.  These curves represent the evolution with collision participants ($N_{part}$) of the expected enhancement factors assuming that a chemical freeze-out of 165 MeV is applicable at all times and in all systems.  The model incorrectly assumes a zero net-baryon density, which is why there is no difference in the curves for both particles and anti-particles.  The volume is calculated as: V = A$^{\alpha}$V$_{0}$, where A = N$_{part}$/2 and V$_{0}$ = 4/3.$\pi$R$^{3}$, where R is the radius of the proton.  As indicated on the figure, the three curves correspond to values of $\alpha$ of 1, 2/3 and 1/3 respectively.  The curve which fits the shape of the data the best is for the case of $\alpha$ = 1/3.  This can be thought of as representing the path length which the participant nucleon ``sees" in the colliding nucleus.  In the case of the $\Xi$, the line goes through all the data points whereas for the $\Lambda$, it has the correct shape but the wrong magnitude, indicating, perhaps, that different particles have different freeze-out temperatures.

Two of the interesting strangeness measurements at low-$p_T$ have been those of radial and elliptic flow ($v_2$).   A hydrodynamically inspired fit to the data shows that the radial flow component of the multi-strange baryons ($\Xi$ and $\Omega$) is lower than that of the more abundant particles ($\pi, K, p, \Lambda$) whilst at the same time, their thermal freeze-out temperature is somewhat higher~\cite{Oldi}.  However, the $v_2$ values of the multi-strange hadrons are consistent with that of the light hadrons.  As $v_2$ is a quantity which integrates over the lifetime of the collision, coupled with the knowledge that they freeze-out thermally earlier than the more abundant particles, the non-zero measurement of the $v_2$ of the multi-strange hadrons can be interpreted as a manifestation of flow in the $partonic$ stage of the collision.

\section{Strangeness at intermediate p$_{T}$}

\subsection{Elliptic Flow}

\begin{figure}[h]
\centering
\mbox{\subfigure[Colour online: $v_2$ as a function of $p_T$ for identified particles.  The right panel represents the same data over a smaller $p_T$ range.]{\includegraphics[width=0.48\textwidth,height=0.39\textwidth]{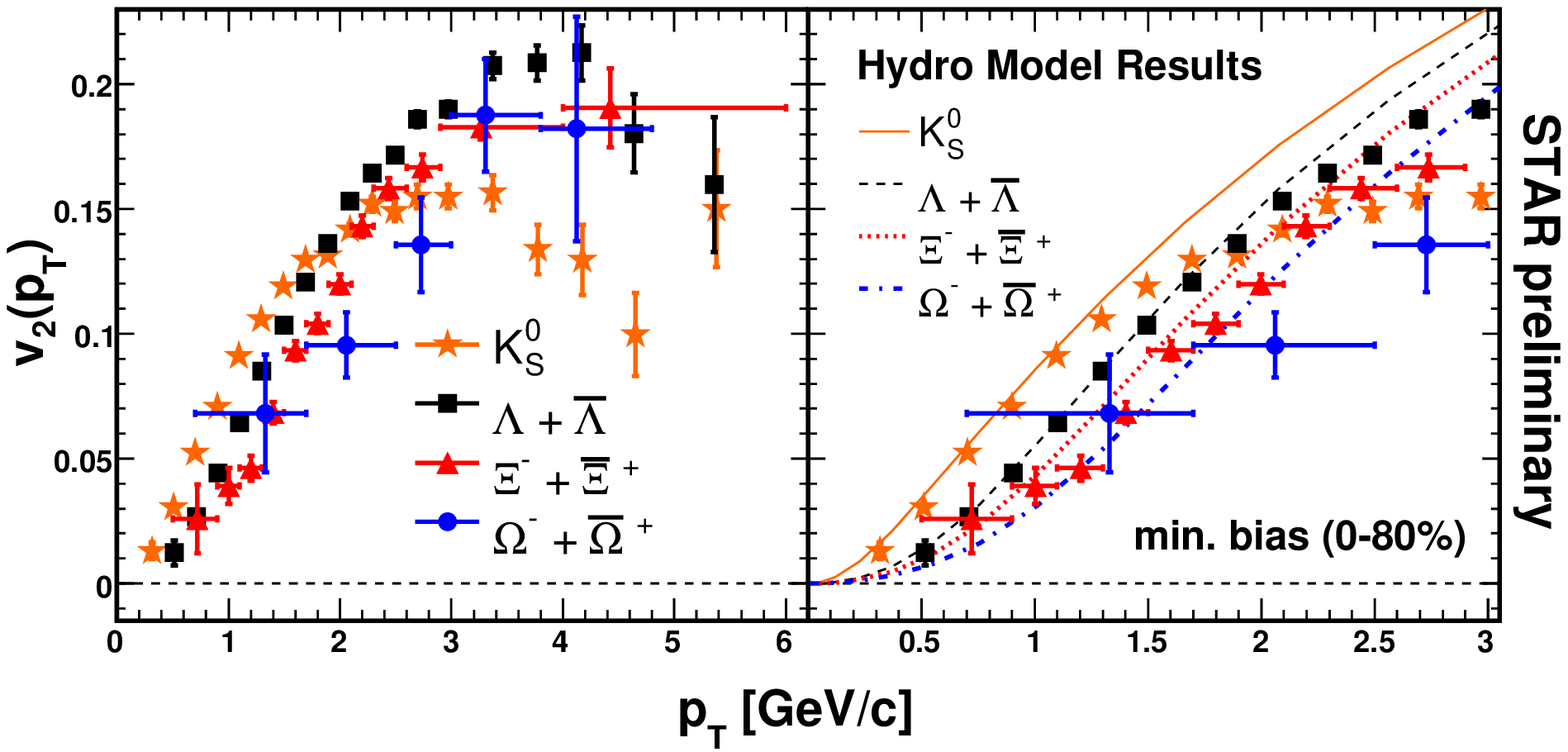}}\quad
\subfigure[Colour online: $v_2$ of identified particles scaled by the number of valence quarks (n), plotted versus $p_T$/n.]{\includegraphics[width=0.45\textwidth,height=0.35\textwidth]{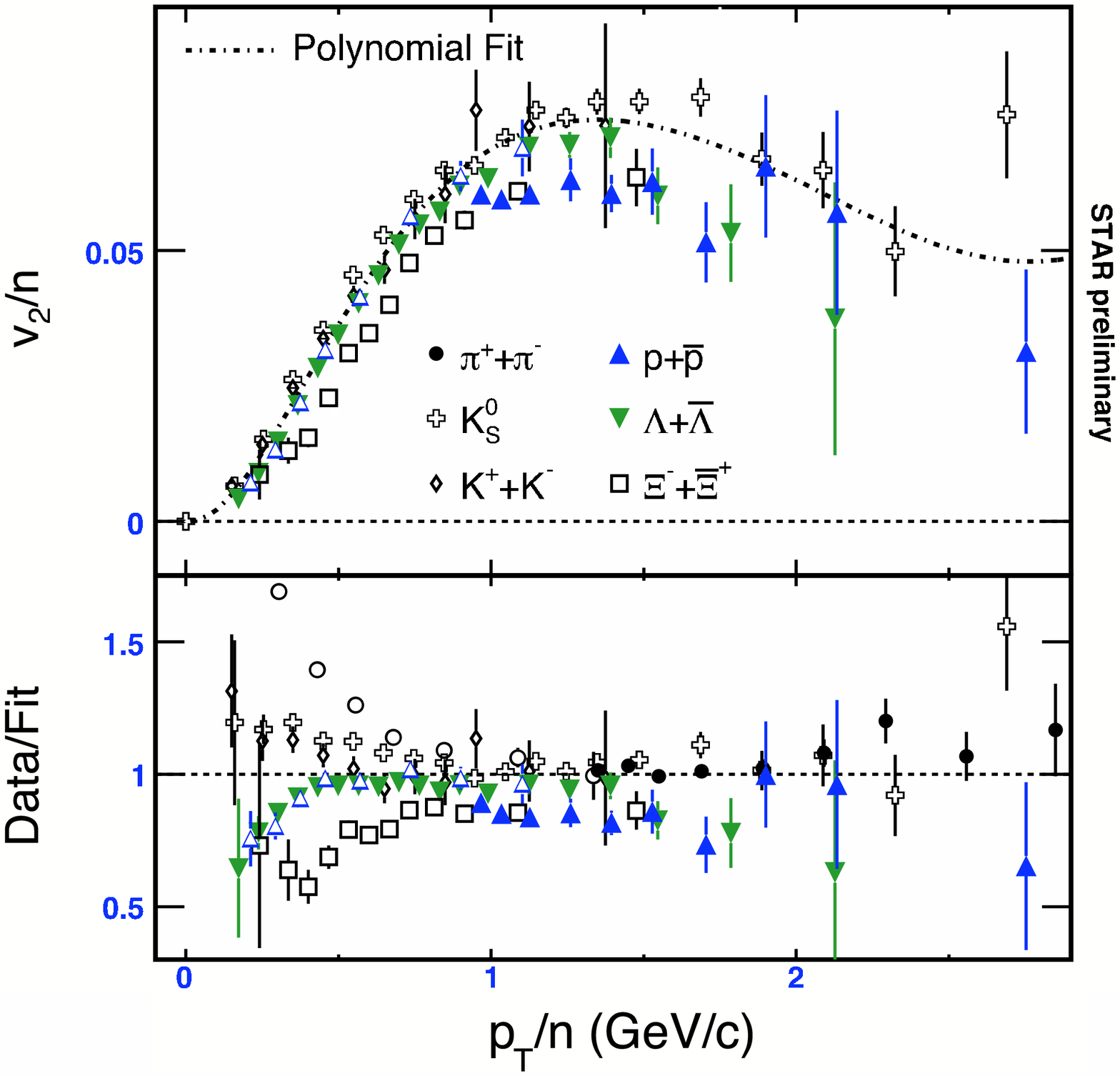}}}
\vspace{-0.5cm}
\caption{}
\label{Fig:v2}
\end{figure}

Although elliptic flow is often thought of as a bulk measurement, STAR has measured $v_2$ out to high-$p_T$.  At lower $p_T$, the $v_2$ for all particle species rises with increasing $p_T$ and is approximately the same, though some mass ordering is observed.  At higher $p_T$ ($>$ 2 GeV/c), as shown in Figure \ref{Fig:v2}a, the $K^0_S$ $v_2$ lies below the $v_2$ of the grouping of baryons.  To exploit this fact, Figure \ref{Fig:v2}b shows the $v_2$ of the hadron, scaled by the number of valence quarks, plotted versus the $p_T$ of the hadron scaled by the same number.  These data are fit by a polynomial and the lower panel shows the deviations of the different particles to the fit.  Although there are some small deviations, above a $p_T/n$ of 1 GeV/c, the fit does a reasonably good job.  This feature can be interpreted as further evidence for partonic collectivity and that $v_2/n$ represents the value of the partonic $v_2$.

\subsection{Particle Ratios}

One of the first indications that particle production at transverse momenta above 2 GeV/c was not purely from modified jet fragmentation came in the form of baryon/meson ratios.  The $p/\pi$ ratio showed an increase to a value consistent with unity for $p_{T} \approx$ 3 GeV/c~\cite{PHENIX_PoverPi}, a value much higher than that observed in elementary collisions~\cite{Elementary_PoverPi}.  The measurement of the baryon/meson ratio was then measured out to higher $p_{T}$ by the STAR experiment~\cite{STAR_LamK0s}, which showed that the ratio reaches a maximum value of about 1.8 at $p_T$ = 2.8 GeV/c, before turning over and decreasing to a value close to that observed in elementary collisions.  The lack of statistics in this data-set mean that a statement on the $p_{T}$ where this comes together cannot be made.

\begin{figure}[htb]
\centering
\mbox{\subfigure[Colour online: The $\Lambda/K^0_S$ ratio as a function of $p_{T}$ for different collision centralities for Au+Au collisions at $\sqrt{s_{NN}}$ = 200 GeV.]{\includegraphics[width=0.49\textwidth,height=0.43\textwidth]{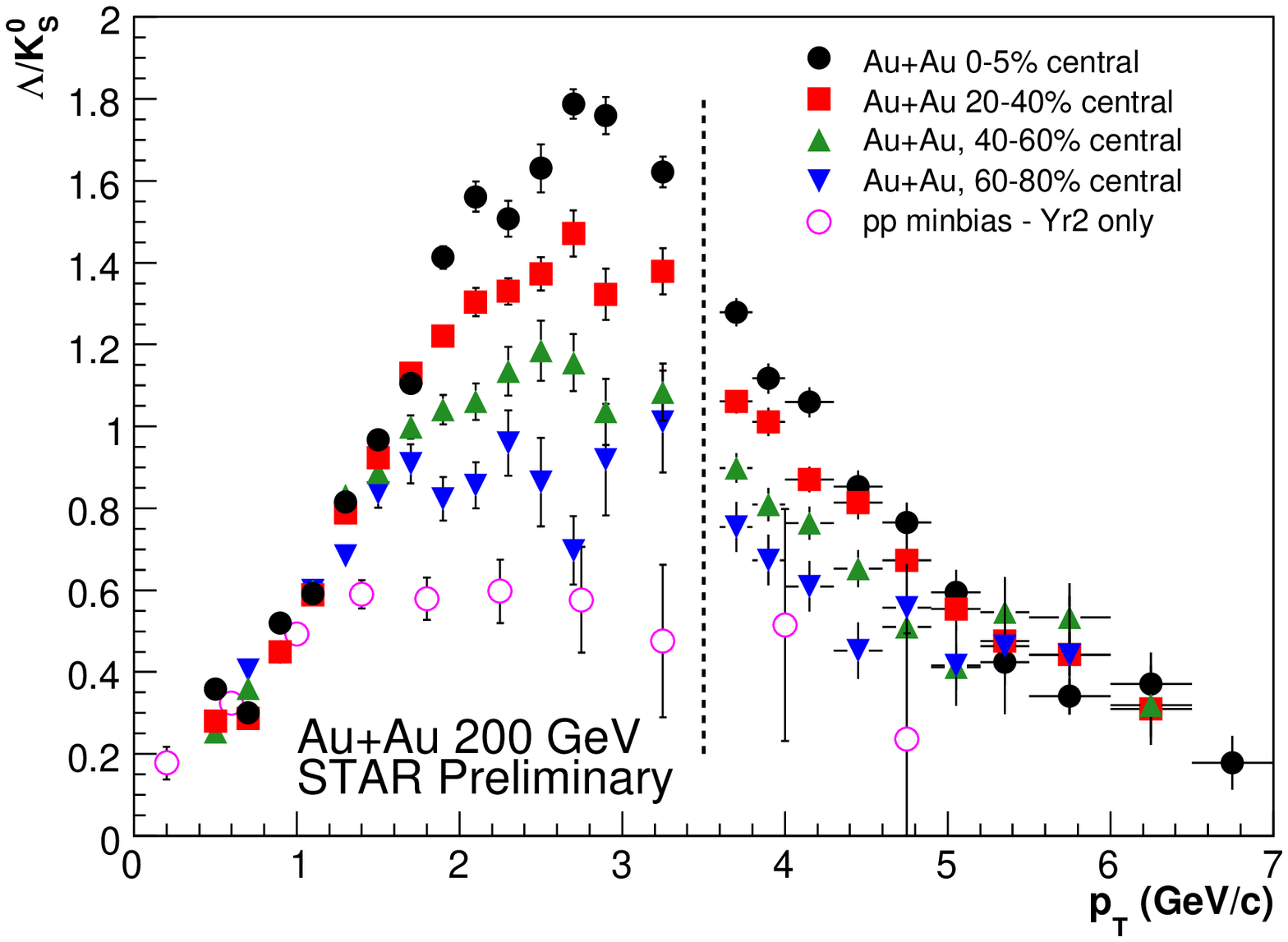}}\quad
\subfigure[Colour online: The $R_{CP}$ ratio measured as a function of $p_{T}$ for different particle species for Au+Au collisions at $\sqrt{s_{NN}}$ = 200 GeV.]{\includegraphics[width=0.49\textwidth,height=0.4\textwidth]{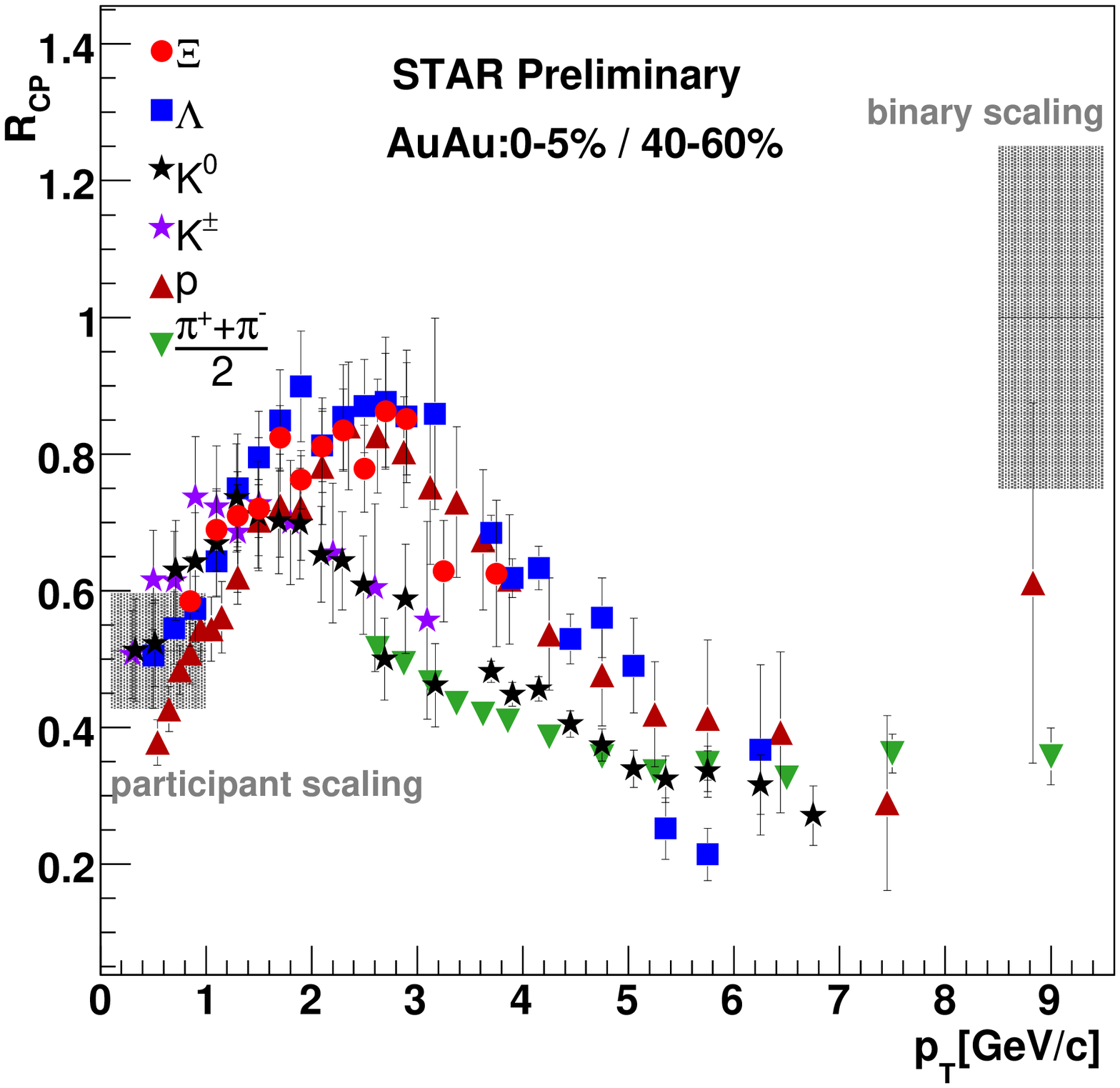}}}
\vspace{-0.5cm}
\caption{}
\label{Fig:LamK0s}
\end{figure}

Figure \ref{Fig:LamK0s}a shows the $\Lambda/K^0_S$ ratio in Au+Au collisions at $\sqrt{s_{NN}}$ = 200 GeV.  The data from Year 2 is plotted to the left of the vertical line at $p_T$ = 3.5 GeV/c, the A+A data to the right of this line represents the extension with the Year 4 data-set.  Whilst the lower statistics data-set extended above a $p_T$ of 3.5 GeV/c, above this point, the statistics are poor.

%

With the extension of the $p_T$ range using the Year 4 data, one can see that by $p_T$ = 6 GeV/c, the $\Lambda/K^0_S$ ratio is similar for all centralities and is consistent with the measurement in the p+p data-set.  Another way of looking at this effect is by measuring the ratio of particle yields in central collisions relative to peripheral collisions, scaled by the relative number of binary collisions.  This ratio is commonly referred to as the $R_{CP}$ and is shown in the right panel of Figure \ref{Fig:LamK0s} for both strange and non-strange particles.  An $R_{CP}$ value of 1 means that particle production simply scales with the number of binary collisions in the system, something which would be expected at high-$p_{T}$ if there are no medium effects.  As observed in the Year 2 data, there is a splitting in the distributions, with the mesons having a smaller $R_{CP}$ than baryons in the range 2 $<$ $p_{T}$ $<$ 6 GeV/c.  Above this value, they show the same suppression.

In this $p_{T}$ range (known as intermediate $p_{T}$), models with novel hadronization mechanisms are required to reproduce these data.  The majority of models used to describe the data use the recombination/coalescence (ReCo) mechanism to form hadrons~\cite{ReCo}.  In these models, heavy $``$constituent$"$ quarks are combined to form hadrons - 3 for baryons and 2 for mesons.  This mechanism naturally leads to a larger baryon-to-meson ratio due to the larger number of quarks at lower $p_{T}$.  In the models, this mechanism is dominant when the parton $p_{T}$ distribution is an exponential, when it behaves as a power-law, then fragmentation dominates.  There are a number of variations of this model which differ in small ways, the models which best describe the data only allow the recombination of thermally produced partons.  Those which allow for recombination between thermal partons and those coming from jets substantially over-predict both the magnitude of the ratio and the turn-over $p_T$.  

\begin{figure}[h!]
\begin{center}
\includegraphics[width=0.5\textwidth, height=0.35\textwidth]{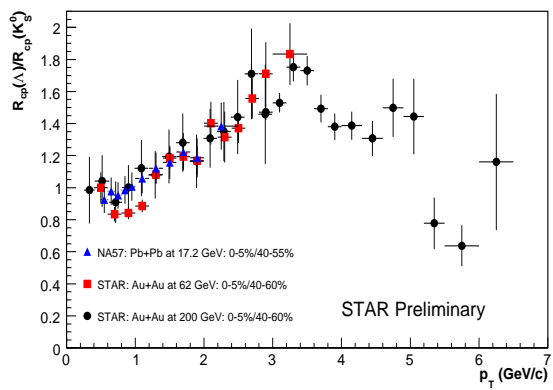}
\end{center}
\caption{Colour online: The $R_{CP}^{\Lambda}/R_{CP}^{K^0_S}$ ratio as a function of $p_{T}$ for three different collision systems.}
\label{Fig:RcpLamK0s}
\end{figure}

One of the interesting topics in strangeness is to investigate in what systems and energies these phenomena are prevalent.  Although the cross-section for high-$p_T$ processes is lower at the CERN-SPS, the $\Lambda/K^0_S$ ratio has been studied out to a $p_T$ of 3 GeV/c~\cite{Schuster}.  This also shows a rise with increasing $p_T$ as has been observed at RHIC, but is at a much larger value due to the larger net-baryon density.  In order to make comparisons between different energies feasible, it was suggested to instead plot the $R_{CP}^{\Lambda}/R_{CP}^{K^0_S}$ ratio which plots the relative difference between the $\Lambda/K^0_S$ ratio in central and peripheral event classes.  This was first suggested elsewhere where Year 2 Au+Au data was plotted on top of the Pb+Pb data at $\sqrt{s_{NN}}$ = 17.2 GeV from NA57 and it showed remarkable agreement~\cite{Dainese}.  In Figure \ref{Fig:RcpLamK0s} we again plot this data, but also add the Year 4 Au+Au data together with data from Au+Au collisions at $\sqrt{s_{NN}}$ = 62 GeV.

As can be seen, there is very good agreement between all three different systems which may be indicative of the same hadroniazation mechanism for all three energies.  The system size dependence of this ratio can be tested once the Cu+Cu data at both $\sqrt{s_{NN}}$ = 200 GeV and $\sqrt{s_{NN}}$ = 62 GeV have been analysed by STAR.

\subsection{Azimuthal Correlations}

\begin{figure}[htb]
\centering
\mbox{\subfigure[Colour online: The correlation functions for the 0-5$\%$ most central events.  The correlations are reflected around $\Delta\phi$ = 0 (open symbols) in order to increase statistics.]{\includegraphics[width=0.49\textwidth,height=0.32\textwidth]{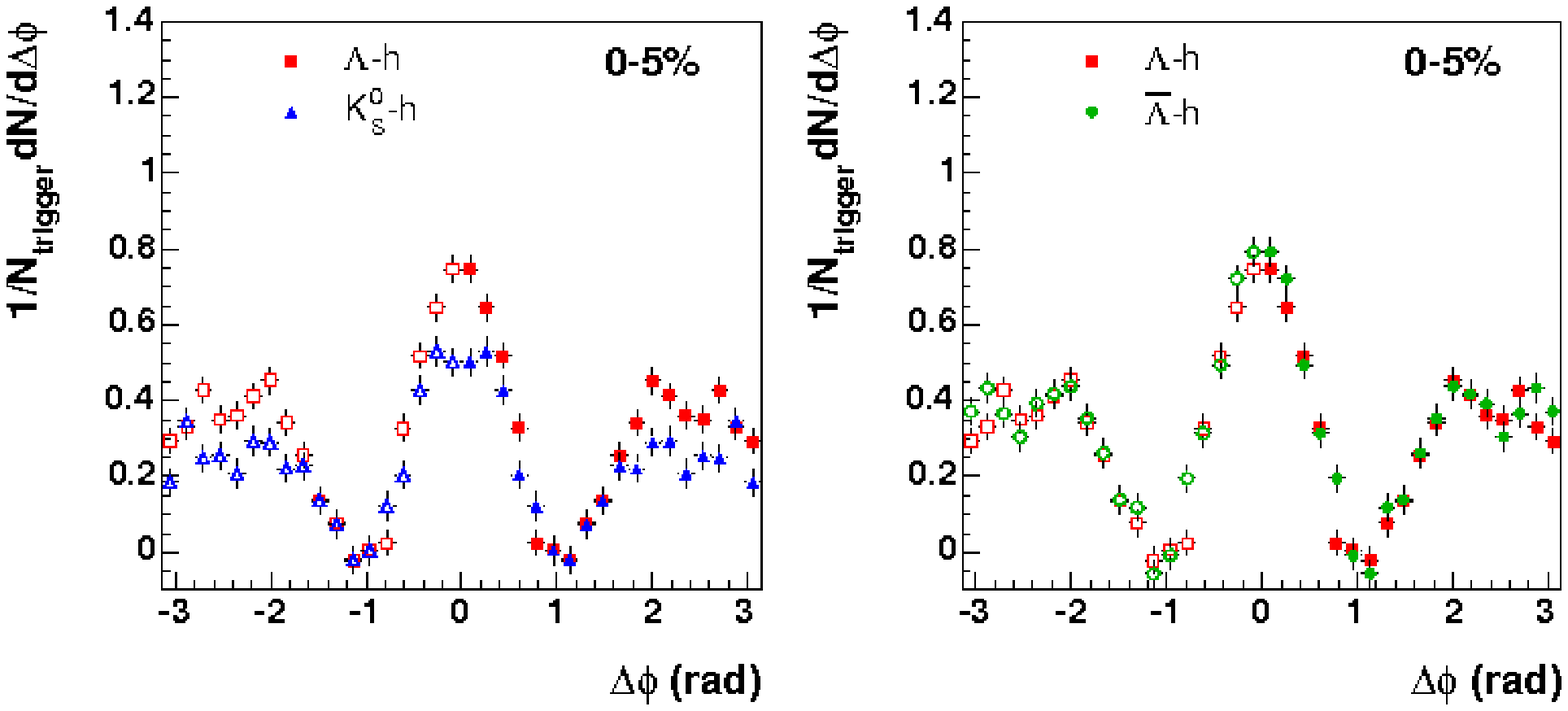}}\quad
\subfigure[Colour online: The correlation functions for the 30-40$\%$ most central events. The correlations are reflected around $\Delta\phi$ = 0 (open symbols) in order to increase statistics.]{\includegraphics[width=0.49\textwidth,height=0.3\textwidth]{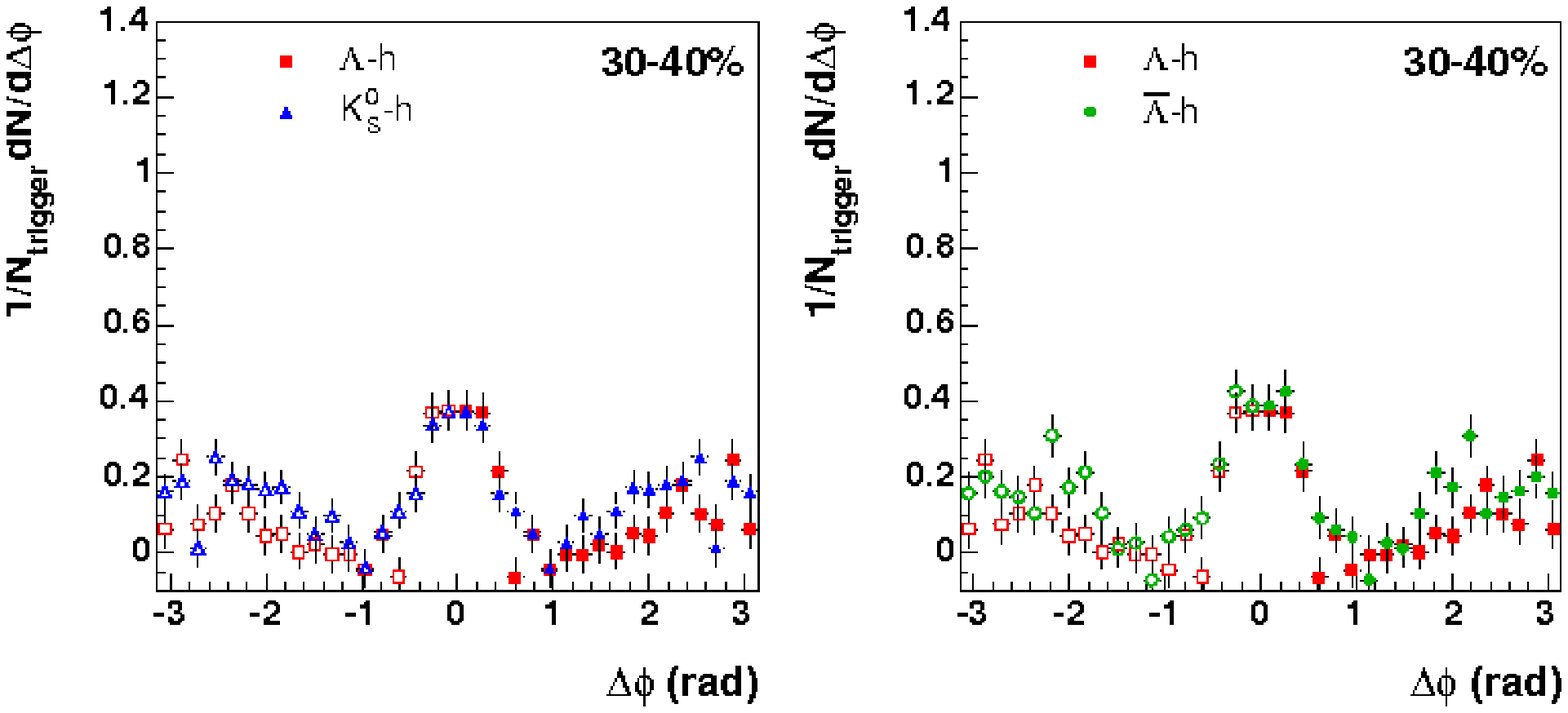}}}
\vspace{-0.5cm}
\caption{}
\label{Fig:RawCorr}
\end{figure}

As well as determining differences between baryons and mesons using particle spectra and ratios, it is also possible to examine these effects using identified two-particle azimuthal correlations.  Two-particle azimuthal correlations have oft been used to substitute for full jet reconstruction in high-multiplicity environments.  STAR has previously shown that in p+p and d+Au collisions, peaks in the correlation are seen for $\Delta\phi$ = 0 and $\Delta\phi$ = $\pi$, indicating the near-side and away-side jets respectively~\cite{STAR_JETS}.  Here, we discuss in detail properties of near-side correlation peaks using identified strange trigger particles.  In Figure \ref{Fig:RawCorr}, we plot example raw correlation functions with $\Lambda$, $\overline{\Lambda}$ and $K^0_S$ trigger particles, associated with an unidentified charged particle, for both the 0-5$\%$ and 30-40$\%$ most central events respectively.  In each case, 3 $<$ $p_T^{Trigger}$ $<$ 3.5 GeV/c, 1 $<$ $p_T^{Assoc}$ $<$ 2 GeV/c and $|\eta|$ $<$ 1.  All correlation functions are normalised to the number of triggers and have been corrected for elliptic flow background, non-uniformities in detector geometry  and reconstruction inefficiency of the associated particles.  The data are fit with two Gaussians (near-side and away-side) on top of a flat background and the associated yield is determined by the area under the Gaussian peak. 

%

\begin{figure}[htb]
\centering
\mbox{\subfigure[Colour online: The yield/trigger for the near-side correlation for d+Au collisions for both charged hadrons ($h-h$) and identified particles.]{\includegraphics[width=0.46\textwidth,height=0.32\textwidth]{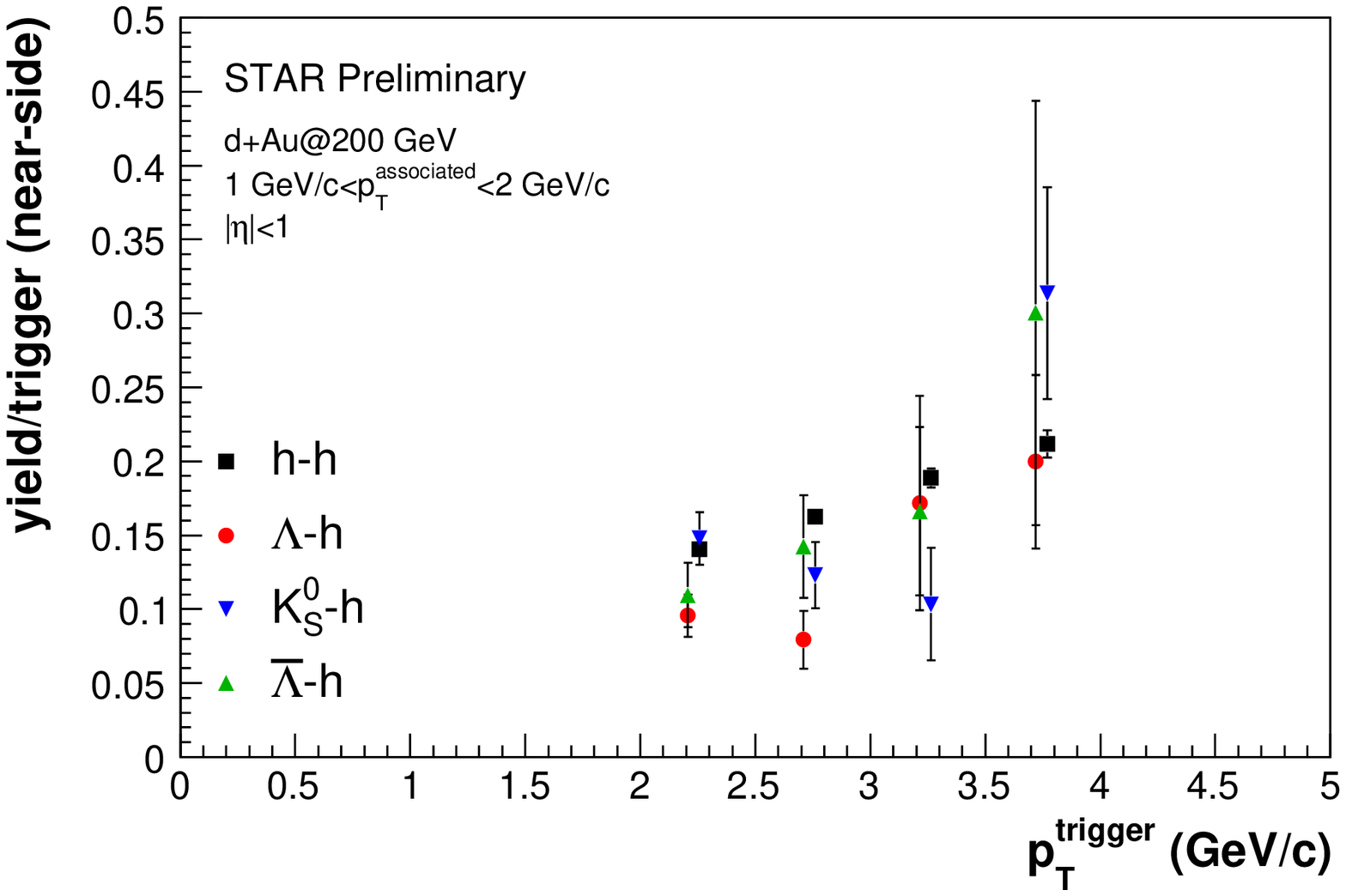}}\quad
\subfigure[Colour online: The yield/trigger for the near-side correlation for Au+Au collisions for both charged hadrons ($h-h$) and identified particles.]{\includegraphics[width=0.46\textwidth,height=0.32\textwidth]{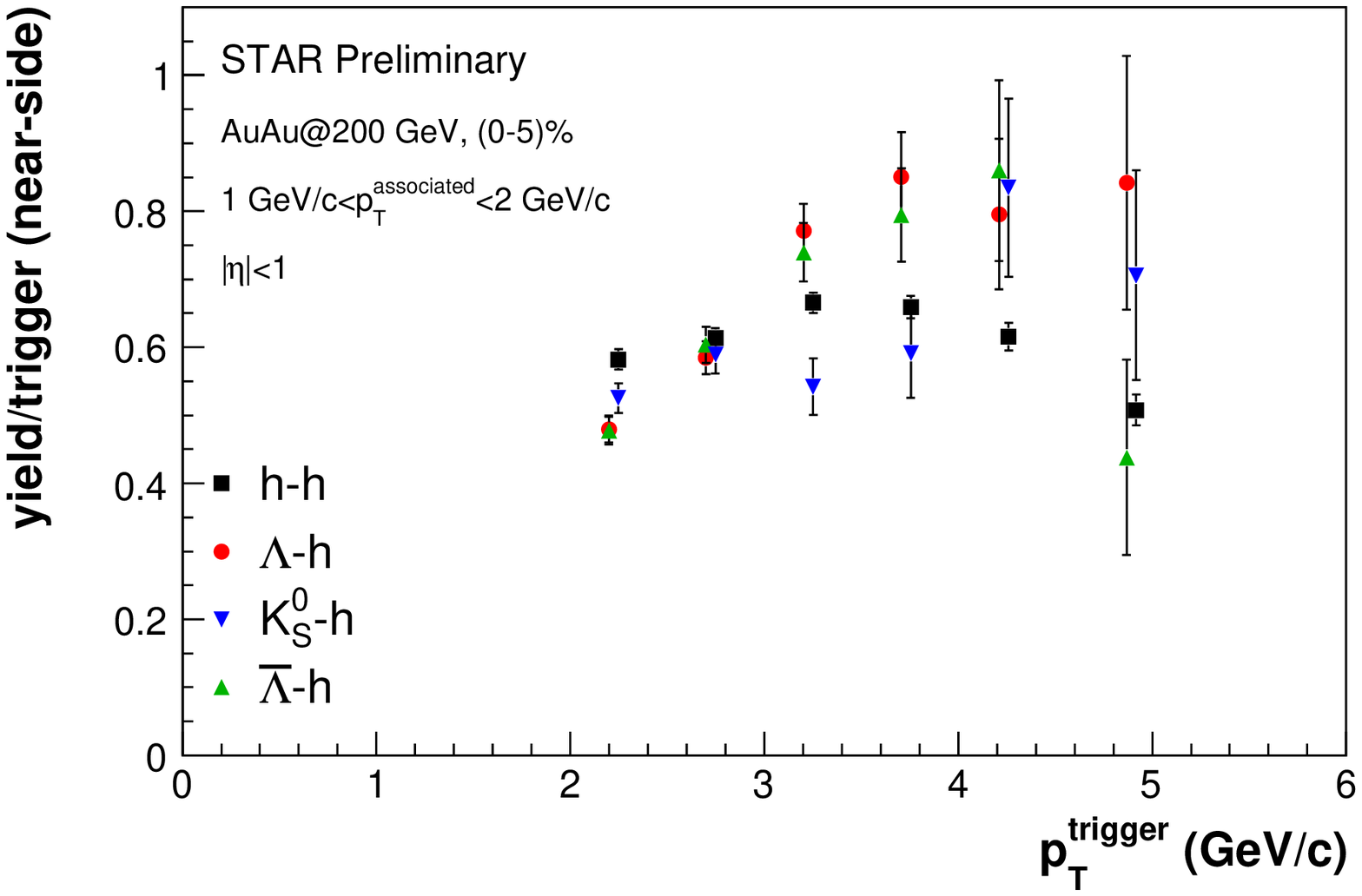}}}
\vspace{-0.5cm}
\caption{}
\label{Fig:NearSideCorr}
\end{figure}

Using these correlation functions, it is possible to measure any differences between baryon- and meson-triggered correlations.  These may be expected if the production of particles comes from different sources (e.g. $\Lambda$ from both $q$ and $g$ jets, $\overline\Lambda$ predominantly from $g$ jets).  Figure \ref{Fig:NearSideCorr} shows the yields/trigger of the near-side correlation for different trigger $p_T$ ranges for d+Au collisions (left panel) and central (0-5$\%$) Au+Au collisions (right panel).  Although there is an increase of 3 to 4 in the near-side yields going from d+Au to Au+Au, in each system, there is no discernible difference, outside the statistical error, for the baryon-, anti-baryon-, or meson-triggered correlations.  This is further evidence to suggest that for the $p_T^{Trigger}$ analysed, particle production is not yet dominated by jet fragmentation. 

%

\begin{figure}[htb]
\centering
\mbox{\subfigure[Colour online: The ratio of the near-side yields in Au+Au collisions for different centralities.]{\includegraphics[width=0.49\textwidth,height=0.32\textwidth]{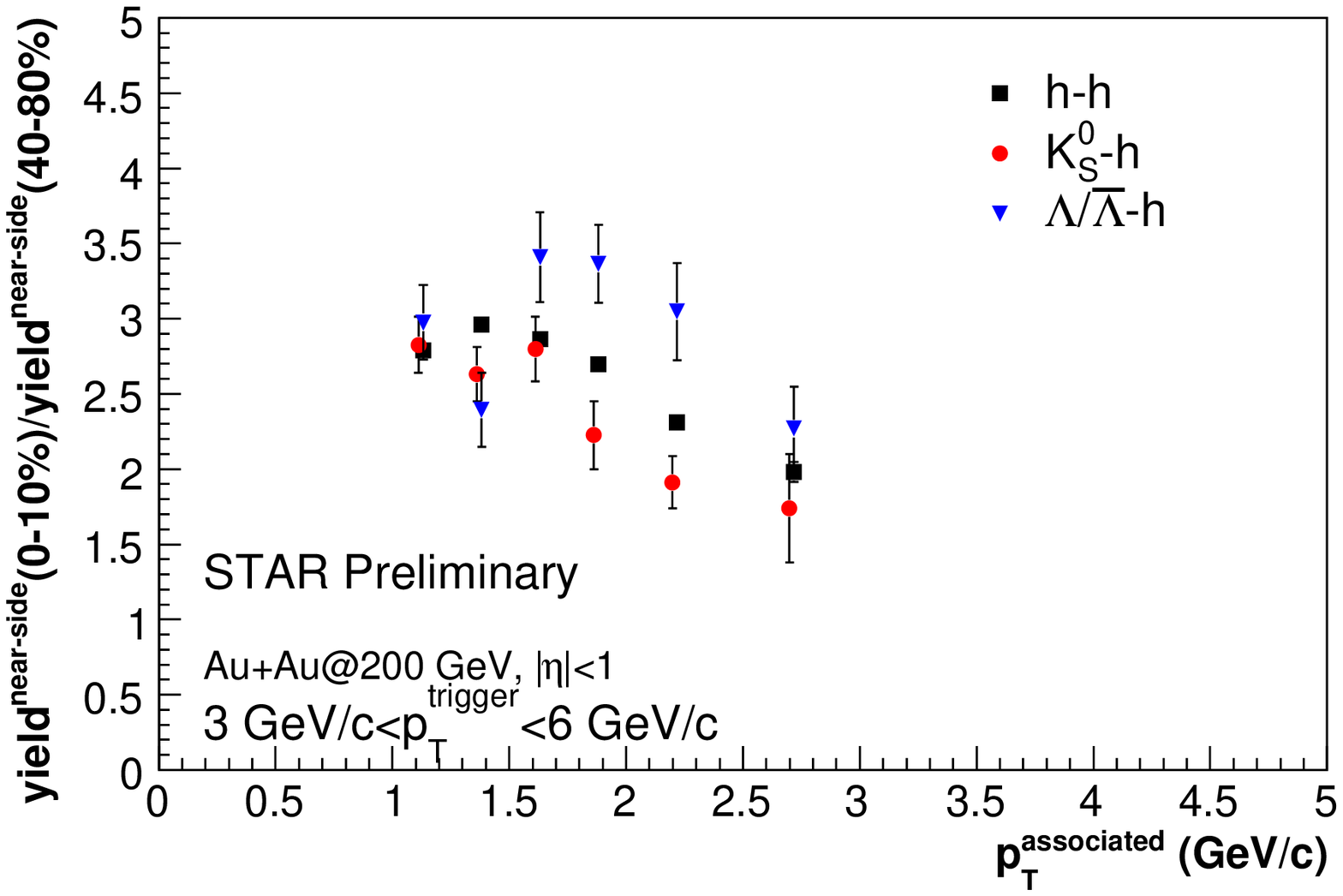}}\quad
\subfigure[Colour online: Predictions for the $R_{CP}$ ratio for different $p_T^{Trigger}$ ranges within a ReCo model.]{\includegraphics[width=0.49\textwidth,height=0.32\textwidth]{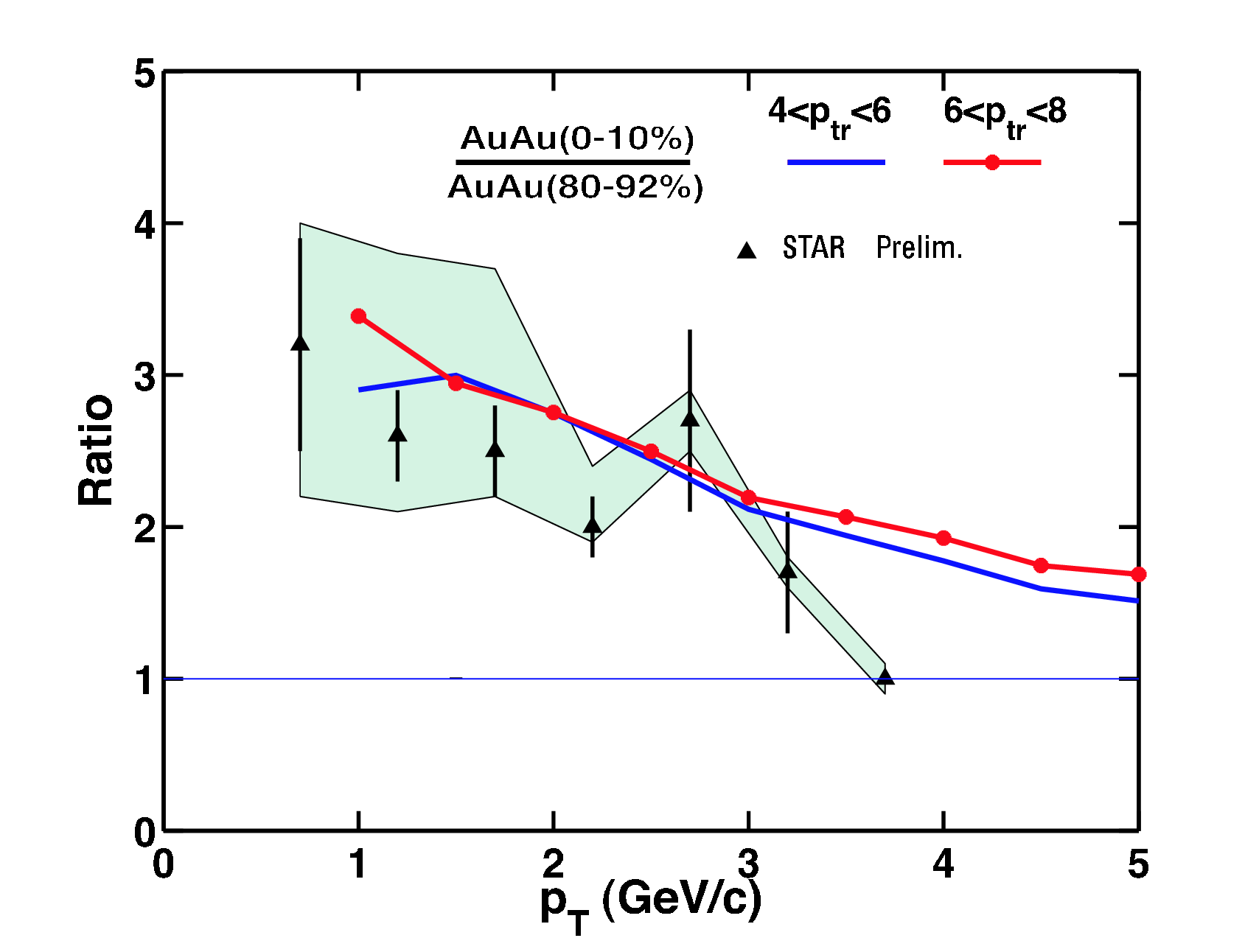}}}
\vspace{-0.5cm}
\caption{}
\label{Fig:RecoCorr}
\end{figure}

If particle production is not coming from fragmentation, then there must be an alternative mechanism to produce azimuthal correlations.  This can be achieved in a ReCo model by allowing the coalescence of a soft ``thermal" parton with that of a hard ``shower" parton (in the case of a meson).  Predictions are shown in Figure \ref{Fig:RecoCorr}b from a ReCo model for the $R_{CP}$ ratio of the near-side yields of unidentified two-particle azimuthal correlations~\cite{RudiJets}.  On the figure, the model is compared to STAR preliminary data from Year 2 (low statistics data-set).

%

Figure \ref{Fig:RecoCorr}a shows the ratio for both unidentified and identified correlations from the high statistics Year 4 data-set and is plotted on the same scale as the predictions in the right panel.  There is a small difference in both the $p_T^{Trigger}$ range as well as the centrality, though this should have little effect on the prediction.  In order to increase statistics, the $\Lambda$ and $\overline{\Lambda}$ correlations have been added together, this is possible as no difference is observed between the separate correlations.  The prediction and the data are in very good agreement in this $p_T^{Trigger}$ range, however this must be analysed more closely.  It has been shown that the $\Delta\phi$ correlations contain a large component of background made up of long range correlations in $\Delta\eta$.  These contributions are $p_T^{Trigger}$ dependent and are most dominant at low $p_T^{Trigger}$ and once taken into account, will lead to a flatter ratio.  These correlations are not present in the model presented here so any agreement could be purely accidental.  These correlations (referred to as a ``ridge$"$) are also probably responsible for the difference in the near-side yields between the Au+Au and d+Au data presented in Figure \ref{Fig:NearSideCorr}.

As well as this ratio, the model has also made predictions regarding $\phi$ and $\Omega$ production which state that contributions from thermal quarks will dominate up to a $p_T$ of 8 GeV/c.  Two consequences of this prediction are that the measured $\Omega/\phi$ ratio should show a linear rise with $p_T$ and that any $\Omega$- or $\phi$-triggered two particle correlations are suppressed and not observed.  Whilst the $\Omega$- and $\phi$-triggered correlations are still being studied, the $\Omega/\phi$ ratio was presented elsewhere at this conference~\cite{SarahBlyth}.  This ratio does rise up to a $p_T$ of 4 GeV/c at which point it starts to turn-over.  The difference in shape and the level of uncertainties on this do not allow for a strong conclusion to be made.  

\section{Summary}

In summary, we have presented recent results on strangeness production at RHIC as measured by the STAR experiment.  By fitting statistical models to the data, it was shown that the enhancement of the strange particles can be reproduced using the assumption that the volume scales with $N_{part}^{1/3}$.  At intermediate-$p_T$, measurements of identified particle $v_2$, the $\Lambda/K^0_S$ ratio and the identified particle $R_{CP}$ all point to models of hadronization other than jet fragmentation.  In identified two-particle azimuthal correlations, we found that there is no dependence on the trigger species in either d+Au or Au+Au collisions and when calculating a ratio of near-side yields, this was in agreement with a recombination model which allows for the recombination of thermal and shower particles.  One crucial test of this model will be correlations with multi-strange particles.  STAR will be able to perform this analysis when the full Year 4 statistics have been analysed.

\section{Acknowledgements}
We thank the RHIC Operations Group and RCF at BNL, and the NERSC Center at LBNL for their support. This work was supported in part by the HENP Divisions of the Office of Science of the U.S. DOE; the U.S. NSF; the BMBF of Germany; IN2P3, RA, RPL, and EMN of France; EPSRC of the United Kingdom; FAPESP of Brazil; the Russian Ministry of Science and Technology; the Ministry of
Education and the NNSFC of China; Grant Agency of the Czech Republic, FOM of the Netherlands, DAE, DST, and CSIR of the Government of India; Swiss NSF; the Polish State Committee for Scientific  Research; and the STAA of Slovakia


\section{References}
\begin{thebibliography}{9}
\bibitem{Rafelski}Rafelski J and Muller B 1982, \PRL $\bf{48}$ 1066.
\bibitem{NA57}Antinori F \etal (NA57 Collaboration) 2006, \JPG $\bf{32}$ 427.
\bibitem{RedlichEnhance} Redlich K and Tounsi A 2002, $Euro. \ Phys. \ J.$ C $\bf{24}$ 589.
\bibitem{NA49}Mitrovski M (for the NA49 Collaboration) these proceedings.
\bibitem{RedlichCurves}Redlich K, private communication.
\bibitem{Oldi}Oldenburg M (for the STAR Collaboration), these proceedings.
\bibitem{PHENIX_PoverPi}Adcox K \etal (PHENIX Collaboration) 2004, \PR C $\bf{69}$ 024904.
\bibitem{Elementary_PoverPi}Abreu P \etal (DELPHI Collaboration) 2000, $Euro. \ Phys. \ J. $ C $\bf{17}$ 207.
\bibitem{STAR_LamK0s}Adams J \etal (STAR Collaboration) 2006, $preprint$ $nucl$$-$$ex/0601042$.
\bibitem{ReCo}Fries R J, Muller B, Nonaka C, Bass S A 2003, \PR C $\bf{68}$ 044902.
\nonum  Greco V, Ko C M and Levai P 2003, \PR C $\bf{68}$, 034904.
\nonum  Hwa R C and Yang C B 2003, \PR C $\bf{67}$ 034902.
\bibitem{Schuster}Schuster T and Laszlo A (for the NA49 Collaboration), these proceedings.
\bibitem{Dainese}Bruno G E and Dainese A (for the NA57 Collaboration) 2005, $preprint$ $nucl$$-$$ex/0511020$.
\bibitem{STAR_JETS}Adler C \etal (STAR Collaboration) 2003, \PRL $\bf{90}$ 082302
\bibitem{RudiJets}Hwa R and Tan Z 2005, \PR C $\bf{72}$ 057902.
\bibitem{SarahBlyth}Blyth S-L (for the STAR Collaboration), these proceedings.

\end{thebibliography}
\end{document}